\documentclass[aps,prd,twocolumn,groupedaddress,showpacs]{revtex4}

\usepackage{graphicx,dcolumn,bm,amssymb,amsmath,latexsym,amsfonts}

\begin{document}

\newcommand{\nonu}{\nonumber}
\newcommand{\sm}{\small}
\newcommand{\noi}{\noindent}
\newcommand{\npg}{\newpage}
\newcommand{\nl}{\newline}
\newcommand{\bc}{\begin{center}}
\newcommand{\ec}{\end{center}}
\newcommand{\be}{\begin{equation}}
\newcommand{\ee}{\end{equation}}
\newcommand{\beal}{\begin{align}}
\newcommand{\eeal}{\end{align}}
\newcommand{\bea}{\begin{eqnarray}}
\newcommand{\eea}{\end{eqnarray}}
\newcommand{\bnabla}{\mbox{\boldmath $\nabla$}}
\newcommand{\univec}{\textbf{a}}
\newcommand{\VectorA}{\textbf{A}}
\newcommand{\Pint}

\title{Generalized black diholes}

\author{I. Cabrera-Munguia$^{1,}$\footnote{cabrera@zarm.uni-bremen.de}, Claus L\"ammerzahl$^{2,}$\footnote{laemmerzahl@zarm.uni-bremen.de}, L. A. L\'opez$^{3,}$\footnote{lalopez@uaeh.edu.mx}
and Alfredo Mac\'{\i}as$^{1,}$\footnote{amac@xanum.uam.mx}}
\affiliation{$^{1}$Departamento de F\'isica, Universidad Aut\'onoma Metropolitana-Iztapalapa A.P. 55-534, M\'exico D.F. 09340, M\'exico\\
$^{2}$ZARM, Universit\"{a}t Bremen, Am Fallturm, D-28359 Bremen, Germany\\
$^{3}$\'Area Acad\'emica de Matem\'aticas y F\'isica., UAEH, carretera Pachuca-Tulancingo km 4.5, C.P. 42184, Pachuca, Hidalgo, M\'exico}


\begin{abstract}
A $5$-parametric exact solution, describing a binary system composed of identical counter-rotating black holes endowed with opposite electromagnetic charges, is constructed. The addition of the angular momentum parameter to the static Emparan-Teo dihole model introduces magnetic charges into this two-body system. The solution can be considered as an extended model for describing ge\-ne\-ra\-li\-zed black diholes as dyons. We derive the explicit functional form of the horizon half-length pa\-ra\-me\-ter $\sigma$ as a function of the Komar parameters: Komar mass $M$, electric/magnetic charge $Q_{E}/Q_{B}$, angular momentum $J$, and a coordinate distance $R$, where the parameters $(M,J,Q_{E},Q_{B},R)$ cha\-rac\-te\-ri\-ze the upper constituent of the system, while $(M,-J,-Q_{E},-Q_{B},R)$ are associated with the lower one. The addition of magnetic charges enhances the standard Smarr mass formula in order to take into account their contribution to the mass. The solution contains, as particular cases, two solutions already discussed in the literature.

\end{abstract}
\pacs{04.20.Jb,04.70.Bw,97.60.Lf}

\maketitle

\section{INTRODUCTION}
Black dihole (BDH) systems have been proposed by Emparan \cite{Emparan, EmparanTeo} as  static binary configurations of identical black holes endowed with opposite electric (or magnetic) charges, which are unbalanced by means of a conical singularity in between \cite{Israel}. These two-body systems carry an electric (or magnetic) dipole moment, and the electromagnetic duality provides the corresponding dual con\-fi\-gu\-ra\-tions.

The addition of an angular momentum parameter ge\-ne\-ra\-lizes these BDH configurations \cite{Manko,ILLM} and means that the system is now composed of a pair of dyons \cite{Schwinger}; i.e., due to rotation of electric charges, the constituents are now endowed with both electric and magnetic dipole moments (monopole electric and magnetic charges).

Tomimatsu proposed in $1984$ \cite{Tomimatsu1}, that due to the magnetic charges in the binary system, the standard Smarr mass formula \cite{Smarr} does not hold. It should be ge\-ne\-ra\-li\-zed to include the contribution of magnetic charges to the mass. Kleihaus \emph{et al.} \cite{Jutta} considered black holes with magnetic monopole or dipole hair, in Einstein-Maxwell theory and some extensions of it, and show that the co\-rres\-pon\-ding black hole solutions satisfy a ge\-ne\-ra\-li\-zed Smarr type mass formula, in agreement with Tomimatsu's proposal.

On one hand, following this idea we considered ex\-pli\-ci\-tly the magnetic charges generated by the rotation of electrically charged black holes. Moreover, we cons\-truc\-ted a $4$-parametric asymptotically flat exact solution in \cite{ILLM}. This generalized stationary Emparan-like solution is endowed with magnetic monopole charges, and electric dipole moment, but it does not contain any magnetic dipole moment.

On the other hand, Manko \emph{et al.} \cite{Manko} introduced a $5$-parametric asymptotically flat exact solution, where ins\-tead of magnetic monopole charges, a magnetic dipole moment produced by the rotation of electrically charged black holes is considered. They enlarge the $4$-parametric Cabrera-Munguia \emph{et al.} solution \cite{ILLM} to a $5$-parametric one, by means of the introduction of a magnetic moment parameter $b$, and they hide the magnetic charges of Cabrera-Munguia \emph{et al.} solution in favor of a magnetic dipole. By setting the magnetic moment $b=0$, the Cabrera-Munguia \emph{et al.} solution is easily recovered. In fact, each solution can be straightforwardly obtained from the other one, i.e., by introducing a magnetic dipole moment parameter (Cabrera-Munguia $\rightarrow$ Manko) or by killing it (Manko $\rightarrow$ Cabrera-Munguia).

Therefore, both solutions are in fact two faces, two particular cases of a more general $5$-parametric exact solution, including magnetic charge and magnetic dipole moment parameters. The electromagnetic potential results to be invariant under the transformation $Q_{E}\leftrightarrow i Q_{B}$. This means that an observer will measure the same electromagnetic effects if we exchange the electric and magnetic potentials. The solution should provide a physical parametrization in terms of the five physical Komar parameters, i.e., the Komar mass $M$, electric/magnetic charges $Q_{E}/Q_{B}$, angular momentum $J$, and coordinate distance $R$. The upper black hole has $(M,J,Q_{E},Q_{B},R)$ while the lower one contains $(M,-J,-Q_{E},-Q_{B},R)$.

The outline of the paper is as follows. In Sec. II, the axis conditions for a $5$-parametric exact solution describing a two-body system of identical counter-rotating Kerr-Newman (KN) black holes with a massless strut in between \cite{Israel} are considered and solved. In Sec. III the explicit form of the horizon half-length parameter $\sigma$ in terms of the five physical Komar parameters $(M,J,Q_{E},Q_{B},R)$ is given. Moreover, in Sec. IV we reduce our more general solution to the two physical descriptions already presented in \cite{ILLM,Manko}. The addition of magnetic charges provides us a more general description of the properties of dyonic BDH \cite{Schwinger}. Additionally, the corresponding Smarr formula and its geometrical components containing the proper contribution of the magnetic charges are displayed. In Sec. V the extreme limit of the solution is obtained. Sec. VI is devoted to the concluding remarks.

\section{$5$-PARAMETRIC CLASS OF SOLUTIONS}
Stationary electrovacuum spacetimes can be described by means of the line element \cite{Papapetrou}
\be ds^{2}=f^{-1}\left[e^{2\gamma}(d\rho^{2}+dz^{2})+\rho^{2}d\varphi^{2}\right]- f(dt-\omega d\varphi)^{2},
\label{Papapetrou}\ee

\noi where $f(\rho,z)$, $\omega(\rho,z)$, and $\gamma(\rho,z)$ are the metric func\-tions which can be calculated through the following system of equations
\bea \begin{split}
f&={\rm{Re}(\cal{E})}+ |\Phi|^{2},\\
\omega_{\rho}&=-\rho f^{-2}{\rm{Im}}( {\cal{E}}_{z}+ 2 \Phi\bar{\Phi}_{z}),\\
\omega_{z}&=\rho f^{-2}{\rm{Im}}( {\cal{E}}_{\rho}+ 2 \Phi\bar{\Phi}_{\rho}),\\
4\gamma_{\rho}&=\rho f^{-2} \left[|{\cal{E}}_{\rho}+
2\bar{\Phi}\Phi_{\rho}|^{2} -|{\cal{E}}_{z}+ 2\bar{\Phi}\Phi_{z}|^{2}\right]\\
&- 4\rho f^{-1}(|\Phi_{\rho}|^{2}- |\Phi_{z}|^{2}),\\
2\gamma_{z}&=\rho f^{-2}{\rm{Re}} \left[({\cal{E}}_{\rho}+
2\bar{\Phi}\Phi_{\rho})(\bar{{\cal{E}}}_{z}+ 2\bar{\Phi}\Phi_{z})\right]\\
&-4\rho f^{-1} {\rm{Re}(\bar{\Phi}_{\rho}\Phi_{z})},
\label{metricfunctions}\end{split}\eea

The set of Eqs. (\ref{metricfunctions}) contains the complex potentials $({\cal{E}}, \Phi)$, which can be determined from the so-called Ernst equations \cite{Ernst},
\bea \begin{split}  \left({\rm{Re}} {\cal{E}}+|\Phi|^{2}\right)\Delta{\cal{E}}&=(\bnabla{\cal{E}}+
2\bar{\Phi}\bnabla \Phi)\bnabla {\cal{E}}, \\
 \left({\rm{Re}}{\cal{E}}+|\Phi|^{2}\right)\Delta \Phi&=(\bnabla{\cal{E}}+
2\bar{\Phi}\bnabla\Phi)\bnabla\Phi, \label{ERNST} \end{split} \eea

\noi where \bnabla\, and $\Delta$ are the gradient and Laplace operators defined in Weyl-Papapetrou cylindrical coordinates $(\rho,z)$. The subscripts $\rho$ and $z$ denote partial differentiation, the bar over a symbol represents complex conjugation and $|x|^{2}=x \bar{x}$. In addition, $\Phi=-A_{4} + i A_{3}^{'}$ is the electromagnetic potential, whose components are the electric potential $A_{4}$ and the potential $A^{'}_{3}$ associated with the magnetic potential $A_{3}$. The metric functions $f$, $\omega$, and $\gamma$ are determined by the Ernst equations (\ref{ERNST}).

Once we know the complex Ernst potentials on the symmetry axis, we can use the Sibgatullin's method (SM), based on the soliton theory, for solving the nonlinear equations (\ref{ERNST}), to obtain straightforwardly the complex Ernst potentials \cite{Sibgatullin,RMJ} for the whole spacetime. For a binary system, the explicit solution for the whole space is obtained by setting $N=2$ in the formulas of the last part of Sec. III of Ref.\cite{RMJ}. Then, the explicit solution contains a set of twelve algebraic parameters $\{\alpha_{n},f_{j},\beta_{j}\}$, for $n=\overline{1,4}$ and $j=1,2$. Due to the presence of a total magnetic charge and NUT sources \cite{NUT}, this $12$-parametric exact solution is not asymptotically flat at spatial infinity. Hence, the \emph{axis conditions} should be established in order to get rid of such monopolar terms.

Therefore, the axis conditions turn out to be very important in order to obtain an asymptotically flat exact solution which describes a two-body system of KN sources (subextreme or hyperextreme sources) with a massless strut in between, i.e., a well-known conical line sin\-gu\-la\-ri\-ty \cite{Israel}. The axis conditions can be reduced to an algebraic system of equations given by \cite{ILLM}
\bea  \begin{split}  {\rm{Im}}[\mathfrak{\bar{a}}_{-}(\mathfrak{g}_{-}+\mathfrak{h}_{-})]= 0, \qquad
{\rm{Im}}[\mathfrak{\bar{a}}_{+}(\mathfrak{g}_{+}+\mathfrak{h}_{+})]=0,&  \\
\mathfrak{g}_{\pm}=\left|
\begin{array}{ccccc}
0 & 2 & 2 & 1\pm1 & 1\pm1 \\
1 & {} & {} & {} & {}  \\
1 & {} & (\mathfrak{a}_{\pm})  \\
0 & {} & {} & {} & {} \\
0 & {} & {} & {} & {} \\
\end{array}
\right|,& \\
\mathfrak{h}_{\pm}=\left|
\begin{array}{ccccc}
0 & 1 & 1 & 1 & 1 \\
1 & {} & {} & {} & {}  \\
1 & {} & (\mathfrak{a}_{\pm})  \\
\bar{e}_{1} & {} & {} & {} & {} \\
\bar{e}_{2} & {} & {} & {} & {} \\
\end{array}
\right|,\\
\mathfrak{a}_{\pm}=\left|\begin{array}{cccc}
\pm \gamma_{11} & \pm \gamma_{12} & \gamma_{13}& \gamma_{14} \\
\pm \gamma_{21} & \pm \gamma_{22} & \gamma_{23}& \gamma_{24} \\
M_{11} & M_{12} & M_{13}& M_{14}\\
M_{21} & M_{22} & M_{23}& M_{24}\\
\end{array}
\right|,& \\
M_{jn}=\left[\bar{e}_{j}+2\bar{f}_{j} f(\alpha_{n})\right](\alpha_{n}-\bar{\beta}_{j})^{-1},& \\ f(\alpha_{n})=\sum_{j=1}^{2} f_{j}\gamma_{jn},\qquad
\gamma_{jn}=(\alpha_{n}-\beta_{j})^{-1},&\\
e_{1}=\frac{2 \prod_{n=1}^{4}(\beta_{1}-\alpha_{n})}
{(\beta_{1}-\beta_{2})(\beta_{1}-\bar{\beta}_{1})(\beta_{1}-\bar{\beta}_{2})}-\sum_{k=1}^{2} \frac{2f_{1}\bar{f}_{k}}{\beta_{1}-\bar{\beta}_{k}},&\\
e_{2}=\frac{2 \prod_{n=1}^{4}(\beta_{2}-\alpha_{n})}
{(\beta_{2}-\beta_{1})(\beta_{2}-\bar{\beta}_{1})(\beta_{2}-\bar{\beta}_{2})}-\sum_{k=1}^{2} \frac{2f_{2}\bar{f}_{k}}{\beta_{2}-\bar{\beta}_{k}}.&
\end{split} \label{algebraicequations}\eea

It is worth mentioning that the algebraic equations (\ref{algebraicequations}) represent a generalization of the axis conditions introduced in \cite{ICM} for vacuum solutions. In order to solve these algebraic equations (\ref{algebraicequations}), we note that the first Simon's multipolar moments \cite{Simon} as the total mass $\mathcal{M}$, total electric charge $\mathcal{Q}$, and total magnetic charge $\mathcal{B}$ of the binary system can be calculated asymptotically from the Ernst potentials on the symmetry axis \cite{ILLM}; they read
\be \beta_{1} + \beta_{2} + \bar{\beta}_{1} + \bar{\beta}_{2}=-2\mathcal{M}, \qquad f_{1}+f_{2}= \mathcal{Q}+ i\mathcal{B}.  \label{Total}\ee

\begin{figure}[ht]
\centering
\includegraphics[width=2.0cm, height=6.0cm]{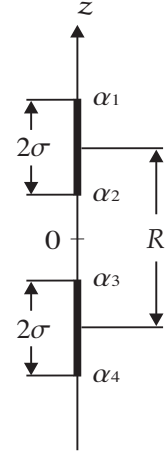}\\
\caption{Two identical KN black holes on the symmetry axis with $\alpha_{1}=-\alpha_{4}=R/2+\sigma$,
$\alpha_{2}=-\alpha_{3}=R/2-\sigma$, and $R>2\sigma$. }
\label{DKNidentical}\end{figure}

By choosing $\beta_{1} + \beta_{2}=-\mathcal{M}:=-2M$, $\mathcal{Q}:=0$, and $\mathcal{B}:=0$, we are describing a system of two identical counter-rotating KN black holes (or relativistic disks) of mass $M$, endowed with opposite electric/magnetic charge $Q_{E}/Q_{B}$ and separated by a supporting strut in between \cite{Israel}. The constant parameters $\alpha_{n}$ fulfill the conditions $\alpha_{1}+\alpha_{4}= \alpha_{2}+\alpha_{3}=0$, as shown in
Fig. \ref{DKNidentical}. They can be written down in terms of the coordinate distance $R$ and the horizon half-length $\sigma$ of each rod describing the black holes as follows:
\be \alpha_{1}=-\alpha_{4}=\frac{R}{2}+\sigma, \qquad \alpha_{2}=-\alpha_{3}=\frac{R}{2}-\sigma.\ee

An explicit solution to the algebraic equations (\ref{algebraicequations}) reads
\bea \begin{split} f_{1,2}&= \pm  \frac{q_{o}+i b_{o}}{\sqrt{p+i\delta}},\qquad
\beta_{1,2}= -M\pm\sqrt{p+i \delta},\\
p&=R^{2}/4-M^{2}+\sigma^{2}, \\
\delta&=\sqrt{(R^{2}-4M^{2})[M^{2}-\sigma^{2}-\mu (Q_{o}^{2}+B_{o}^{2})]},\\
q_{o}&:=Q_{o}(R/2-M), \qquad b_{o}:=B_{o}(R/2-M), \\
\mu&:=\frac{R-2M}{R+2M}. \label{implicitsolution2}\end{split}\eea

Since the identical KN black holes are counter-rotating and have opposite electric charges, the full metric exhibits an equatorial antisymmetry property in the sense proposed by Ernst \emph{et al.} \cite{EMR} and further studied by Sod-Hoffs \emph{et al.} \cite{Jordi}. The solution Eq. (\ref{implicitsolution2}) is reported by Manko \emph{et al.} in Ref. \cite{Manko} as an extension of the one introduced by Cabrera-Munguia \emph{et al.} in \cite{ILLM}. It is worthwhile to stress the fact that a suitable parametrization can give us straightforward information not only for the identical case under consideration but also for the unequal case \cite{ICM,Varzugin,Varzugin1,AB}.

By using Eq. (\ref{implicitsolution2}), one is able to prove that the Ernst potentials on the upper part of the symmetry axis read
\bea \begin{split}
e(z)&=\frac{e_{+}}{e_{-}},\qquad f(z)=\frac{2(q_{o}+i b_{o}) }{e_{-}}, \\
e_{\pm}&=z^{2} \mp 2M z + 2M^{2}-R^{2}/4-\sigma^{2}-i\delta.\end{split}\eea

The constant parameters $q_{o}$ and $b_{o}$ are associated with the electric and magnetic dipole moment, respectively. One should notice that the transformation $\sigma \rightarrow i \sigma$ in Eq. (\ref{implicitsolution2}) leads to a description of relativistic disks (hyperextreme sources). Nevertheless, in what follows in this paper we are mainly interested in the description of a $5$-parametric asymptotically flat exact solution describing a binary system composed by identical KN black holes. The black holes will be characterized by the physical Komar parameters $\{M, J, Q_{E}, Q_{B}\}$ and the coordinate distance $R$. The Ernst potentials and metric functions for the whole space are obtained by means of the SM. They read
\begin{widetext}
\bea \begin{split}
{\cal{E}}&=\frac{\Lambda-\Gamma}{\Lambda+\Gamma},\quad \Phi=\frac{\chi}{\Lambda+\Gamma},\quad
f=\frac{|\Lambda|^{2}-|\Gamma|^{2}+|\chi|^{2}}{|\Lambda+\Gamma|^{2}}, \quad \omega=\frac{{\rm{Im}}\left[(\Lambda+\Gamma)\bar{\mathcal{G}}-\chi\bar{\mathcal{I}}\right]}{|\Lambda|^{2}-
|\Gamma|^{2}+|\chi|^{2}},\quad e^{2\gamma}=\frac{|\Lambda|^{2}-|\Gamma|^{2}+|\chi|^{2}}{\kappa_{o}^{2} r_{1}r_{2}r_{3}r_{4}},\\
\Lambda&=4\sigma^{2}[\kappa_{+}+2(q_{o}^{2}+b_{o}^{2})](r_{1}-r_{3})(r_{2}-r_{4})+ R^{2}[\kappa_{-}-2(q_{o}^{2}+b_{o}^{2})](r_{1}-r_{2})(r_{3}-r_{4})\\
&+ 2\sigma R(R^{2}-4\sigma^{2})
\left[\sigma R(r_{1}r_{4}+r_{2}r_{3})+i\delta(r_{1}r_{4}-r_{2}r_{3})\right],\\
\Gamma&=2M\sigma R(R^{2}-4\sigma^{2})[ \sigma R(r_{1}+r_{2}+r_{3}+r_{4})
-(2M^{2}-i\delta)(r_{1}-r_{2}-r_{3}+r_{4})],\\
\chi&=-4(q_{o}+i b_{o})\sigma R[(R-2\sigma)(\epsilon_{+}+4M^{2})(r_{1}-r_{4})+(R+2\sigma)(\epsilon_{-}-4M^{2})(r_{2}-r_{3})], \\
\mathcal{G}&=-2 z \Gamma + 2\sigma R[ 4\sigma \kappa_{+}(r_{1}r_{2}-r_{3}r_{4})  +2R\kappa_{-}(r_{1}r_{3}-r_{2}r_{4})-M(R-2\sigma)\nu_{+}(r_{1}-r_{4})
-M(R+2\sigma)\nu_{-}(r_{2}-r_{3})  ], \\
\mathcal{I}&= -(q_{o}+i b_{o})\{4M [2\sigma^{2}(R^{2}-4M^{2}-2i\delta)(r_{1}r_{2}+r_{3}r_{4})
+ R^{2}(2M^{2}-2\sigma^{2}+i\delta) (r_{1}r_{3}+r_{2}r_{4})]-2(R^{2}-4\sigma^{2})\\
&\times\left[ 2M\left[(\epsilon_{+}+4M^{2})r_{1}r_{4}-(\epsilon_{-}-4M^{2})r_{2}r_{3}\right]
+ \sigma R\left[(\epsilon_{+}+8M^{2})(r_{1}+r_{4})+(\epsilon_{-}-8M^{2})(r_{2}+r_{3})+8\sigma M R\right]\right]\}, \\
\kappa_{o}&:=4\sigma^{2}R^{2}(R^{2}-4\sigma^{2}),\quad \kappa_{\pm}:=M^{2}(R^{2}-4\sigma^{2})\pm 2(q_{o}^{2}+b_{o}^{2}), \quad \nu_{\pm}:=\epsilon_{\pm}(R\pm2\sigma)^{2}\pm 8(q_{o}^{2}+b_{o}^{2}),\\ \epsilon_{\pm}&:=\sigma R\mp(2M^{2}-i \delta),  \label{5-parametros}  \end{split} \eea
\end{widetext}

\noi where $r_{n}$ are given by
\bea \begin{split} r_{1,2}&=\sqrt{\rho^{2}+\left(z-R/2 \mp \sigma\right)^{2}}, \\
r_{3,4}&=\sqrt{\rho^{2}+\left(z+R/2 \mp \sigma\right)^{2}}. \end{split}\eea

\section{PHYSICAL PARAMETRIZATION AND LIMITS OF THE SOLUTION}
In order to write $\sigma$ in terms of physical Komar pa\-ra\-me\-ters \cite{Komar}, $M$, $J$, $Q_{E}$, $Q_{B}$, and the coordinate distance $R$, we will apply the well-known Tomimatsu's formulas \cite{Tomimatsu1} to the upper object, since the black holes are identical,
\bea \begin{split} M&= -\frac{1}{8\pi}\int_{H} \omega \Psi_{z}\, d\varphi dz, \\
Q_{E}&=\frac{1}{4\pi}\int_{H}\omega A_{3z}^{'}\, d\varphi dz, \,\, Q_{B}=\frac{1}{4\pi}\int_{H}\omega A_{4z} \, d\varphi dz, \\
J&=-\frac{1}{8\pi}\int_{H}\omega\left[1+\frac{1}{2}\omega \Psi_{z}
-\tilde{A}_{3}A_{3z}^{'}-(A_{3}^{'}A_{3})_{z}\right]d\varphi dz , \end{split}\label{Tomy}\eea

\noi with $\tilde{A}_{3}:=A_{3}+ \omega A_{4}$ and $\Psi={\rm Im}(\cal{E})$. The magnetic potential $A_{3}$ is the real part of the Kinnersley's potential $\Phi_{2}$ \cite{Kinnersley}. By means of the SM \cite{RMJ} it can be written as follows:
\be A_{3}={\rm Re}\left(\Phi_{2}\right)={\rm Re}\left(-i\frac{I}{E_{-}}\right)=-z A_{3}^{'} + {\rm Im} \left(  \frac{\mathcal{I}}{\Lambda+\Gamma}\right).\ee

The upper black hole horizon is defined as a null hypersurface $H=\{-\sigma\leq z - \frac{R}{2}\leq \sigma,\,0 \leq \varphi \leq 2\pi,\, \rho\rightarrow 0\}$. Thus, $M$ represents the individual mass of each black hole source. Moreover, the electric and magnetic charges read
\bea \begin{split}
Q_{E}&=\frac{Q_{o}(R^{2}-4M^{2})+2B_{o}\delta}
{R^{2}-4\sigma^{2}-4\mu (Q_{o}^{2}+B_{o}^{2})},\\
Q_{B}&=\frac{B_{o}(R^{2}-4M^{2})-2Q_{o}\delta}{
R^{2}-4\sigma^{2}-4\mu (Q_{o}^{2}+B_{o}^{2})}.\label{charges}\end{split}\eea

Combining both Eqs. (\ref{charges}) one gets
\be |Q_{E}^{2}+Q_{B}^{2}|= \frac{(Q_{o}^{2}+B_{o}^{2})(R^{2}-4M^{2})}{R^{2}-4\sigma^{2}-4\mu(Q_{o}^{2}+B_{o}^{2})},\ee

\noi which suggests that we introduce a new auxiliary variable $X$ as follows:
\be X:=\frac{Q_{o}^{2}+B_{o}^{2}}{|Q_{E}^{2}+Q_{B}^{2}|} \label{theX}.\ee

Hence, $\sigma$ can be written as a function of this auxiliary variable as follows,
\be \sigma=\sqrt{X[M^{2}-|Q_{E}^{2}+Q_{B}^{2}|\mu]+\frac{R^{2}}{4}\left(1-X\right)}, \label{sigmaX} \ee

\noi where $Q_{o}$ and $B_{o}$ can be rewritten as
\bea \begin{split}
Q_{o}&=Q_{E}-Q_{B}\sqrt{X-1},\\
B_{o}&=Q_{B}+Q_{E}\sqrt{X-1}.
\label{dipolemoments}\end{split}\eea

The values of $Q_{o}$ and $B_{o}$ depend on the election of sign made in $Q_{E}$ and $Q_{B}$. On the other hand, following Tomimatsu \cite{Tomimatsu1}, the mass formula reads
\bea \begin{split}
 M &= \frac{\kappa S}{4\pi} +2 \Omega J +\Phi^{H}_{E} Q_{E}+ M_{A}^{S}\\
&=\sigma +2 \Omega J +\Phi^{H}_{E} Q_{E}+ M_{A}^{S}, \label{newSmarr} \end{split}\eea

\noi with $M_{A}^{S}$ an extra boundary term associated with the magnetic charge, which is given by
\be M_{A}^{S}=-\frac{1}{4\pi}\int_{H} \left(A_{3}A_{3}^{'}\right)_{z}\, d\varphi dz. \label{MAS}\ee

The angular velocity $\Omega:=1/\omega^{H}$, where $\omega^{H}$ is the metric function $\omega$ evaluated at the horizon. Moreover, $\Phi^{H}_{E}=-A_{4}^{H}-\Omega A_{3}^{H}$ is the electric potential in the frame rotating with the black hole. Using Eq. (\ref{sigmaX}) and Eq. (\ref{dipolemoments}), a simple calculation leads to the following expressions for $M_{A}^{S}$, $\Phi^{H}_{E}$, and $\Omega$:
\bea \begin{split}
M_{A}^{S}&=Q_{B}(Q_{B}\phi^{H}-Q_{E}\Omega),\qquad \Phi^{H}_{E}= Q_{E}\phi^{H}-Q_{B}\Omega,\\
\Omega&=\frac{\mu}{2}\,\frac{(R+2\sigma)\sqrt{X-1}}{M[R+2\sigma-(R-2M)X]-\mu |Q_{E}^{2}+Q_{B}^{2}|X}, \\
\phi^{H}&:=\frac{\mu}{2}\, \frac{R+2\sigma-(R-2M)X}{M[R+2\sigma-(R-2M)X]-\mu |Q_{E}^{2}+Q_{B}^{2}|X}.
\label{newgeometrical} \end{split}\eea

As Tomimatsu proposed \cite{Tomimatsu1}, if the potential $A_{3}$ does not vanish at the two ends of the horizon $H$, the term $M_{A}^{S}$ does not disappear and the Smarr mass formula must take into account the contribution of the magnetic charge $Q_{B}$ to the mass. Combining Eqs.(\ref{newgeometrical}) with each other, it is easy to find the enhanced Smarr formula for the mass \cite{Tomimatsu1,ILLM},
\bea \begin{split}
M&=\sigma +\Omega \left[2J-Q_{E}Q_{B}\left(1-\frac{Q_{B}^{2}}{Q_{E}^{2}}\right) \right] \\ &+\Phi^{H}_{E}\left(1+\frac{Q_{B}^{2}}{Q_{E}^{2}}\right)Q_{E}
=\sigma +2\Omega (J-Q_{E}Q_{B})\\
&+\Phi^{H}_{EL} Q_{E}+\Phi^{H}_{MAG}Q_{B},
\label{SmarrFormula} \end{split}\eea

\noi where
\be \Phi^{H}_{EL}=Q_{E}\phi^{H},\quad \Phi^{H}_{MAG}=Q_{B}\phi^{H}.\label{Potelecmag} \ee

Replacing $\sigma$ from Eq. (\ref{sigmaX}) into the enhanced Smarr formula Eq. (\ref{SmarrFormula}) leads us to the following result
\be X=1 + \frac{4(J-Q_{E}Q_{B})^{2}}{[M(R+2M)+|Q_{E}^{2}+Q_{B}^{2}|]^{2}};\label{laX}\ee

\noi thus, the explicit form of $\sigma$ in terms of physical Komar parameters reads
\begin{widetext}
\be \sigma= \sqrt{M^{2}-\left[ |Q_{E}^{2}+Q_{B}^{2}|+\frac{\mathcal{J}^{2}[(R+2M)^{2}+4|Q_{E}^{2}+Q_{B}^{2}|]}
{[M(R+2M)+|Q_{E}^{2}+Q_{B}^{2}|]^{2}}\right]\frac{R-2M}{R+2M}},  \quad \mathcal{J}:=J-Q_{E}Q_{B}.\label{sigma}\ee
\end{widetext}

Notice that the angular momentum presents an additional contribution from the electromagnetic charges, in agreement with Tomimatsu \cite{Tomimatsu1}. Eq. (\ref{sigma}) for $\sigma$ in terms of the five physical parameters is one of the main results of our paper. Another important result is the straightforward reduction of this solution, Eqs. (\ref{5-parametros}), to the two particular solutions presented by Cabrera-Munguia \emph{et al.} in \cite{ILLM}, and by Manko \emph{et al.} in \cite{Manko}.

As we shall see in the next section, a correct introduction, in the mass formula, of the boundary term $M_{A}^{S}$ gives us a proper contribution of the magnetic charge $Q_{B}$ to the physical and geometrical properties of the system.

\vspace{-0.3cm}\subsection{Physical and Geometrical properties}
Replacing Eq. (\ref{laX}) into Eq. ({\ref{dipolemoments}), it is straightforward to obtain explicit formulas for the electric and magnetic dipole moments,
\bea \begin{split}
2q_{o}&=\left[Q_{E}-\frac{2Q_{B}\mathcal{J}}{M(R+2M)+ |Q_{E}^{2}+Q_{B}^{2}|}\right](R-2M),\\
2b_{o}&=\left[Q_{B}+\frac{2Q_{E}\mathcal{J}}{M(R+2M)+ |Q_{E}^{2}+Q_{B}^{2}|}\right](R-2M).
\label{dipoles} \end{split}\eea

Since $|Q_{E}^{2}+Q_{B}^{2}|$ remains always positive, one notes from Eqs.(\ref{dipoles}) that the term $q_{o}+ib_{o}$ remains invariant under the transformation $Q_{E}\leftrightarrow iQ_{B}$. This means that one observer will measure the same electromagnetic effects if one exchanges the electric and magnetic potentials.

On the other hand, the surface gravity $\kappa$ and area of the horizon $S$ can be obtained directly from
Eq. (\ref{5-parametros}) and without any previous knowledge of the explicit form of $\sigma$. In order to calculate $\kappa$, one uses the formula \cite{Tomimatsu1},
\be \kappa=\sqrt{-\Omega^{2}e^{-2\gamma^{H}}},\ee

\noi where $\gamma^{H}$ is the metric function $\gamma$ evaluated at the horizon. A straightforward calculation leads us to the following expressions for the surface gravity and the area of the horizon:
\bea \begin{split}
\kappa&=\frac{R \sigma(R+2\sigma)}{2M(M+\sigma)(R+2\sigma)(R+2M)-Q^{2}(R-2M)^{2}},\\
S&=4\pi \left[ 2M(M+\sigma)\left(1+\frac{2M}{R}\right)- \frac{Q^{2}(R-2M)^{2}}{R(R+2\sigma)}\right],\\ Q^{2}&:=|Q_{E}^{2}+Q_{B}^{2}|X, \end{split}\eea

\noi with $X$ given by Eq. (\ref{theX}). The energy-momentum tensor associated with the strut gives us the interaction force between the black holes \cite{Israel,Weinstein},
\be \mathcal{F}=\frac{1}{4}(e^{-\gamma_{0}}-1)= \frac{M^{2}}{R^{2}-4M^{2}}+  \frac{ |Q_{E}^{2}+Q_{B}^{2}|\mu R^{2}}{(R^{2}-4M^{2})^{2}},\label{force} \ee

\noi where $\gamma_{0}$ is the value of the metric function $\gamma$ on the region of the strut. One should notice that the strut between the KN black holes disappears in the limit $R\rightarrow \infty$, and the bodies are isolated. In this limit Eq. (\ref{sigma}) reduces to $\sigma=\sqrt{M^{2}-|Q_{E}^{2}+Q_{B}^{2}|-\mathcal{J}^{2}/M^{2}}$ and the electric and magnetic dipole moments behave as $q_{o}\sim Q_{E}R/2$ and $b_{o}\sim Q_{B}R/2$ respectively. Finally, if $R\rightarrow 2M$, the two horizons overlap each other, both angular velocities stop and the system evolves as one single Schwarzschild black hole.

\section{Two Particular Cases}
\vspace{-0.3cm}\subsection{The case with $B_{o}=0$}
The first particular case of this more general solution Eq. (\ref{5-parametros}) is the Cabrera-Munguia solution \cite{ILLM}. We noticed already that for a vanishing magnetic dipole moment term ($b_{o}=0$), one obtains the following cubic equation:
\be (X-1)\left[X-2\left(1-\frac{2M^{2}}{Q_{E}^{2}(1-\mu)}\right)\right]^{2}-\frac{4J^{2}}{Q_{E}^{4}}=0, \label{cubic}\ee

\noi whose explicit real root solution is given by
\bea \begin{split}
X&=1+ \frac{[a+ [b-a^{3} +\sqrt{b(b-2a^{3})}]^{1/3}]^{2}}{[b-a^{3} +\sqrt{b(b-2a^{3})}]^{1/3}},\\
a&:=\frac{1}{3}\left(1-\frac{4M^{2}}{Q_{E}^{2}(1-\mu)}\right),\quad b:=\frac{2J^{2}}{Q_{E}^{4}},\quad b\geq2a^{3}. \label{solutionX} \end{split}\eea

From Eq. (\ref{dipolemoments}) the monopole magnetic charge reads
\be Q_{B}=-Q_{E}\sqrt{X-1}. \label{magneticcharge}\ee

The functional form of $\sigma$ reduces to \cite{ILLM}:
\be \sigma=\sqrt{X(M^{2}-Q_{E}^{2}\mu X)+\frac{R^{2}}{4}\left(1-X\right)}, \label{sigmaX2} \ee

\noi where the explicit value of $X$ is given by Eq. (\ref{solutionX}). Therefore, the interaction force
Eq. (\ref{force}) now contains a spin-spin interaction,
\be \mathcal{F}= \frac{M^{2}}{R^{2}-4M^{2}}+  \frac{ Q_{E}^{2}\mu R^{2}}{(R^{2}-4M^{2})^{2}}X.\label{force2} \ee

\noi The behavior of the magnetic charges arising from the rotation of electrically charged bodies in a weak electromagnetic field or with slow rotation is already discussed in \cite{ILLM}. To conclude the subsection, it should be pointed out that Eq. (\ref{cubic}) can be also obtained from the mass formula Eq. (\ref{SmarrFormula}).

\vspace{-0.3cm}\subsection{The case with $Q_{B}=0$}
A second particular case of Eq. (\ref{5-parametros}) is the Manko \emph{et al}. \cite{Manko} solution. In this case, $Q_{B}=0$, the electric and magnetic dipole moments read
\bea \begin{split}
2q_{o}&=Q_{E}(R-2M),\\
2b_{o}&=\frac{2Q_{E} J (R-2M)}{M(R+2M)+Q_{E}^{2}}.
\label{dipolesManko} \end{split}\eea

The magnetic dipole moment arises as a consequence of the rotation of electrically charged black holes. Nevertheless, the electric dipole moment $2q_{o}$ does not contain any contribution from the rotation parameter $J$. Hence, it remains electrostatic. This is due mainly to the fact that the rotation effects are associated with the monopole magnetic charge [see Eq. (\ref{dipoles})]. The interaction force in this case remains electrostatic [see Eq. (\ref{force})].

On the other hand, the explicit formula for the horizon $\sigma$ presented in \cite{Manko} can be obtained from
Eq. (\ref{sigma}) by setting $Q_{B}=0$, i.e.,
\be \sigma= \sqrt{M^{2}-\left[Q_{E}^{2}+\frac{J^{2}[(R+2M)^{2}+4Q_{E}^{2}]}
{[M(R+2M)+Q_{E}^{2}]^{2}}\right]\frac{R-2M}{R+2M}}.\label{sigmawithoutQB}\ee

\section{Two-body extreme black holes system}
By setting $\sigma=0$ in Eq. (\ref{5-parametros}), the $4$-parametric extreme solution is obtained. In this limit, the angular momentum parameter reads
\bea \begin{split}
|\mathcal{J}|&= \{ M(R+2M)+|Q_{E}^{2}+Q_{B}^{2}| \}\\
&\times \sqrt{\frac{M^{2}(R+2M)-|Q_{E}^{2}+Q_{B}|^{2}(R-2M)}{(R-2M)[(R+2M)^{2}+4|Q_{E}^{2}+Q_{B}^{2}|]}},
\label{angularmomentumextreme} \end{split}\eea

\noi whose asymptotic expansion leads to the condition
\bea \begin{split} &\frac{|\mathcal{J}|}{M \sqrt{M^{2}-|Q_{E}^{2}+Q_{B}^{2}|}}\simeq 1\\
+&\frac{2M^{4}+|Q_{E}^{2}+Q_{B}^{2}|(M^{2}-|Q_{E}^{2}+Q_{B}^{2}|)}{M (M^{2}-|Q_{E}^{2}+Q_{B}^{2}|)}\left(\frac{1}{R}\right)>1. \end{split} \eea

\noi The inequality $\mathcal{J}^{2}/M^{2}>M^{2}-|Q_{E}^{2}+Q_{B}^{2}|>0$ holds for positive values of the distance
$R \gg 2M$. The equality $\mathcal{J}^{2}/M^{2}=M^{2}-|Q_{E}^{2}+Q_{B}^{2}|$ is reached as the distance grows large enough, tending to infinity; therefore, both black holes are isolated. A careful use of l'H\^{o}pital's rule leads to the extreme limit of the solution Eq. (\ref{5-parametros}):
\begin{widetext}
\bea \begin{split} {\cal{E}}&=\frac{\Lambda-2\alpha M x\Gamma_{+}}{\Lambda+2\alpha M x\Gamma_{+}},\quad \Phi=\frac{2(q_{o}+i b_{o})y\Gamma_{-}}{\Lambda+2\alpha M x\Gamma_{+}},\quad f=\frac{D}{N}, \quad
\omega=\frac{4\alpha^{2}\delta_{o}\,y(x^{2}-1)(y^{2}-1)W}{D}, \quad
e^{2\gamma}=\frac{D}{\alpha^{8}(x^{2}-y^{2})^{4}}, \\
\Lambda&= \alpha^{2}(\alpha^{2}-M^{2})(x^{2}-y^{2})^{2}+ \alpha^{2}M^{2}(x^{4}-1)
+ (q_{o}^{2}+b_{o}^{2})(1-y^{4}) +2i\alpha^{2} \delta_{o}(x^{2}+y^{2}-2 x^{2} y^{2}),\\
\Gamma_{\pm}&= \left(\sqrt{M^{2}-\mu |Q_{E}^{2}+Q_{B}^{2}|X}\mp i\sqrt{\alpha^{2}-M^{2}}\right)
\left[\sqrt{M^{2}-\mu |Q_{E}^{2}+Q_{B}^{2}|X}(x^{2}-1) \pm i\sqrt{\alpha^{2}-M^{2}}(x^{2}-y^{2})\right]\\
&+ \mu |Q_{E}^{2}+Q_{B}^{2}|X(x^{2}-1), \\
D&= [\alpha^{2}(\alpha^{2}-M^{2})(x^{2}-y^{2})^{2}+\alpha^{2}M^{2}(x^{2}-1)^{2} -(q_{o}^{2}+b_{o}^{2})(y^{2}-1)^{2}]^{2}-16\alpha^{4}\delta_{o}^{2}x^{2} y^{2}(x^{2}-1)(1-y^{2}), \\
N&=\{ \alpha^{2}(\alpha^{2}-M^{2})(x^{2}-y^{2})^{2}+ \alpha^{2}M^{2}(x^{4}-1)+(q_{o}^{2}+b_{o}^{2})(1-y^{4})
+ 2\alpha M x[(\alpha^{2}-M^{2})(x^{2}-y^{2})+M^{2}(x^{2}-1)]\}^{2}\\
&+4\alpha^{2}\delta_{o}^{2}\left[\alpha(x^{2}+y^{2}-2x^{2}y^{2})+M x(1-y^{2})\right]^{2}, \\
W&= M\alpha^{2} [(\alpha^{2}-M^{2})(x^{2}-y^{2})(3x^{2}+y^{2})+ M^{2}(3x^{4}+6x^{2}-1) +8\alpha M x^{3}]
+ (q_{o}^{2}+b_{o}^{2})[M(y^{2}-1)^{2}-4\alpha x y^{2}], \\
\delta_{o}&:=\sqrt{(\alpha^{2}-M^{2})[M^{2}-\mu |Q_{E}^{2}+Q_{B}^{2}|X]},\qquad X:=1+ \frac{\mu^{-1}M^{2}-|Q_{E}^{2}+Q_{B}^{2}|}{(\alpha+M)^{2}+|Q_{E}^{2}+Q_{B}^{2}|}, \qquad \alpha:=\frac{R}{2}, \label{extreme}\end{split}\eea
\end{widetext}
\noi where $(x,y)$ are prolate spheroidal coordinates,
\be x=\frac{r_{+}+r_{-}}{2\alpha}, \quad y=\frac{r_{+}-r_{-}}{2\alpha}, \quad r_{\pm}=\sqrt{\rho^{2} +(z\pm\alpha)^{2}}, \label{cilindricas}\ee

\noi related to the cylindrical coordinates $(\rho,z)$ via the relations
\be \rho=\alpha\sqrt{(x^{2}-1)(1-y^{2})}, \qquad z=\alpha xy. \label{cilindricas} \ee

We note that the metric Eq. (\ref{extreme}) fulfills the axis condition for all the regions on the symmetry axis: $\omega(y=\pm1)=0$ for $|z|>\alpha$ and $\omega(x=1)=0$ for $|z|<\alpha$. The Emparan's BDH solution \cite{Emparan,Varzugin1} is obtained from Eq.( \ref{extreme}) if $Q_{B}=0$ and $J=0$. The vacuum solution is obtained for $Q_{E}=Q_{B}=0$ \cite{MRRS,KCH,Varzugin}.

\section{CONCLUDING REMARKS}
In this work, we study the consequences of the addition of an angular momentum parameter $J$ to the static Emparan's BDH models. Therefore, the system is now composed of a pair of dyons \cite{Schwinger}. Due to rotation of electric charges, the KN black holes are now endowed with both electric and magnetic monopole charges (electric and magnetic dipole moments). We construct a $5$-parametric $(M,J,Q_{E},Q_{B},R)$ [$(M,J,q_{o},b_{o},R)$] a\-symp\-to\-ti\-ca\-lly flat exact solution. Our generalized black dihole model reduces, for $b_{o}=0$, to the Cabrera-Munguia \emph{et al.} solution \cite{ILLM} and for $Q_{B}=0$ reduces to the Manko \emph{et al.} solution \cite{Manko}.

The parametrization of the solution in terms of magnetic monopole charges $Q_{B}$ allows a deeper understanding of the physical properties of the spacetime of such configurations. The Smarr mass formula should be enhanced in order to take into account their contributions to the mass, in agreement with Tomimatsu \cite{Tomimatsu1}. Additionally, instead of duality properties, the electromagnetic field remains invariant under the exchange of electric and magnetic potentials, i.e., $Q_{E}\leftrightarrow i Q_{B}$. The rotation induces additional contributions, arising from the magnetic and electric charges, to the permanent electric and magnetic dipole moments \cite{Kleihaus,Volkov}.

On the other hand, we derive the corresponding formula of $\sigma$ in terms of the physical Komar parameters and the coordinate distance. Moreover, since the mass $M$ and the angular momentum $J$, now contain contributions from the gravitational and electromagnetic fields \cite{Carter}, one should expect that the explicit formula of $\sigma$,
Eq. (\ref{sigma}), can give us explicit values for these components. Indeed, we apply the Tomimatsu's formulas,
Eqs. (\ref{Tomy}), for the mass and angular momentum in the following representation \cite{Tomimatsu1,Carter}:
\bea \begin{split} M&=M_{G}+M_{E},\quad J=J_{G}+J_{E}, \\
M_{G}&=-\frac{1}{8\pi}\int_{H}\omega [\Psi_{z}-2 {\rm {Im}}(\Phi \bar{\Phi}_{z})] d\varphi dz, \\M_{E}&=-\frac{1}{4\pi}\int_{H}\omega {\rm {Im}}(\Phi \bar{\Phi}_{z})d\varphi dz,\\
J_{G}&= -\frac{1}{8\pi}\int_{H}\omega\left[1+\frac{1}{2}\omega \Psi_{z}-\omega{\rm {Im}}(\Phi \bar{\Phi}_{z})\right]d\varphi dz,\\
J_{E}&=\frac{1}{4\pi}\int_{H}\omega A_{3}A_{3z}^{'} d\varphi dz, \end{split}\label{Tomi2}\eea

\noi where the subscripts $G$ and $E$ denote the gravitational and electromagnetic components, respectively. Therefore, the gravitational and electromagnetic masses read
\bea \begin{split} M_{G}&=\sigma +2\Omega J_{G},\\
M_{E}&=2\Omega(J_{E}-Q_{E}Q_{B})+Q_{E}\Phi_{EL}^{H}+Q_{B}\Phi_{MAG}^{H}, \end{split}\label{Gravitaelectromag}\eea

\noi where $J_{E}=Q_{E}A_{3}^{H}$. Table \ref{table1} shows a set of numerical values for the five physical parameters of our solution, Eq. (\ref{5-parametros}); the mass and angular momentum are written in terms of their gravitational and electromagnetic components. We noticed in Fig. \ref{SLS1} that the presence of negative mass in the solution generates ring singularities off the axis and can change the sign of the angular momentum parameter. Moreover, the presence of the electric and magnetic charges locates such singularity outside the ergosurface.
\vspace{-0.3cm}\begin{table}[ht]
\centering
\caption{Numerical values showing a decomposition of the mass and angular momentum into their gravitational and electromagnetic components.}
\begin{tabular}{c c c c c c c c  }
\hline \hline
$\sigma$ &$ Q_{E}$&$ Q_{B}$ & $R$ & $M_{G}$ & $M_{E}$ &$J_{G}$ &$J_{E}$\\ \hline
  0.6 & -0.3  & -0.4 & 3.813  & 0.955  & 0.045   & 1.363  & 0.137  \\
  0.6 & -0.2  & -0.1 & 3.813  & -0.579 & -0.421  & -0.285 & -0.06 \\
  1.2 & -0.5  & 0    & 4.871  & 1.267  & 0.033   & 0.957  & 0.043\\
  1.2 & -0.13 & 0    & 4.871  & -0.594 & -0.706  & -0.226 & -0.086\\
  0   & -0.3  & 0.2  & 2.4    & 0.962  & 0.038   & 3.243  & 0.046\\
  0   & -0.1  & 0.1  & 2.4    & -0.879 & -0.121  & -0.206 & -0.011\\
  0   & -0.2  & -0.3 & 2.6    & 0.015  & -1.015  & 0.017  & 0.032\\
  \hline \hline
\end{tabular}
\label{table1}
\end{table}

Since the positive mass theorem \cite{SchoenYau1,SchoenYau2} establishes that a regular solution contains a total positive ADM mass \cite{ADM}, then $M>0$. Nevertheless, the condition $M>0$ is not enough to ensure regularity of the solution. Hence, we need to be sure that the denominator of the Ernst potentials is free of zeros. The numerical analysis depicted by Table \ref{table1} reveals in Fig. \ref{SLS1} that if the individual Komar masses are positive, our solution does not develop ring singularities off the axis.
\begin{figure}[ht]
\begin{minipage}{0.49\linewidth}
\centering
\includegraphics[width=4.25cm,height=5cm]{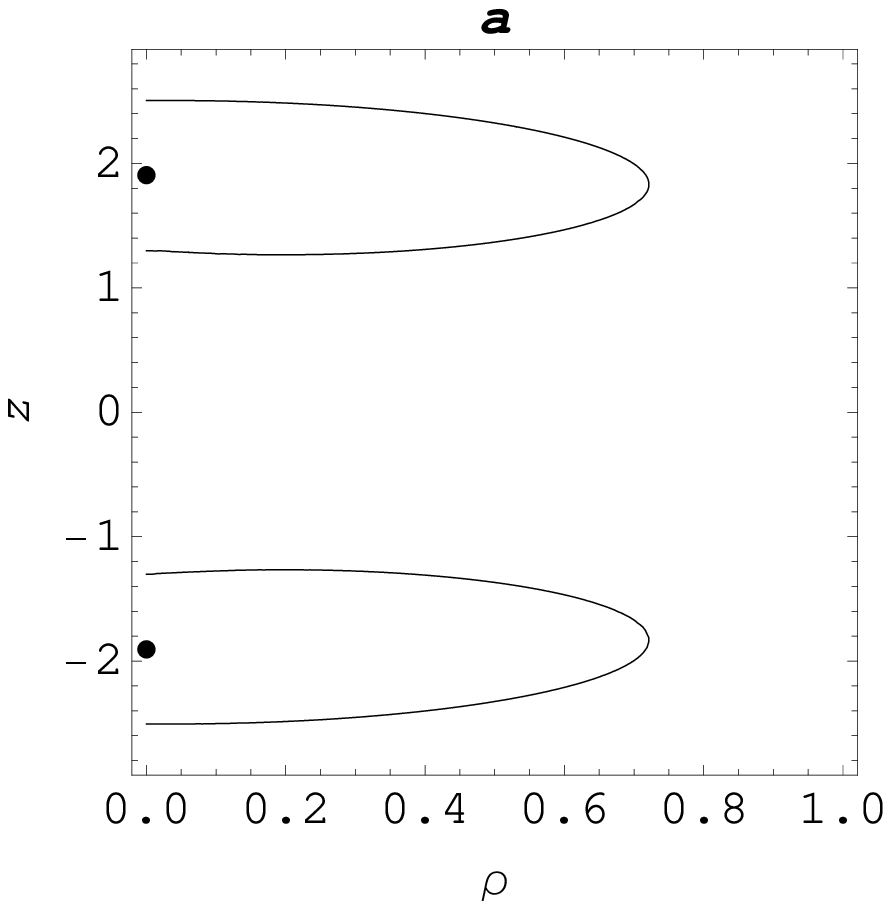}
\end{minipage}
\begin{minipage}{0.49\linewidth}
\centering
\includegraphics[width=4.25cm,height=5cm]{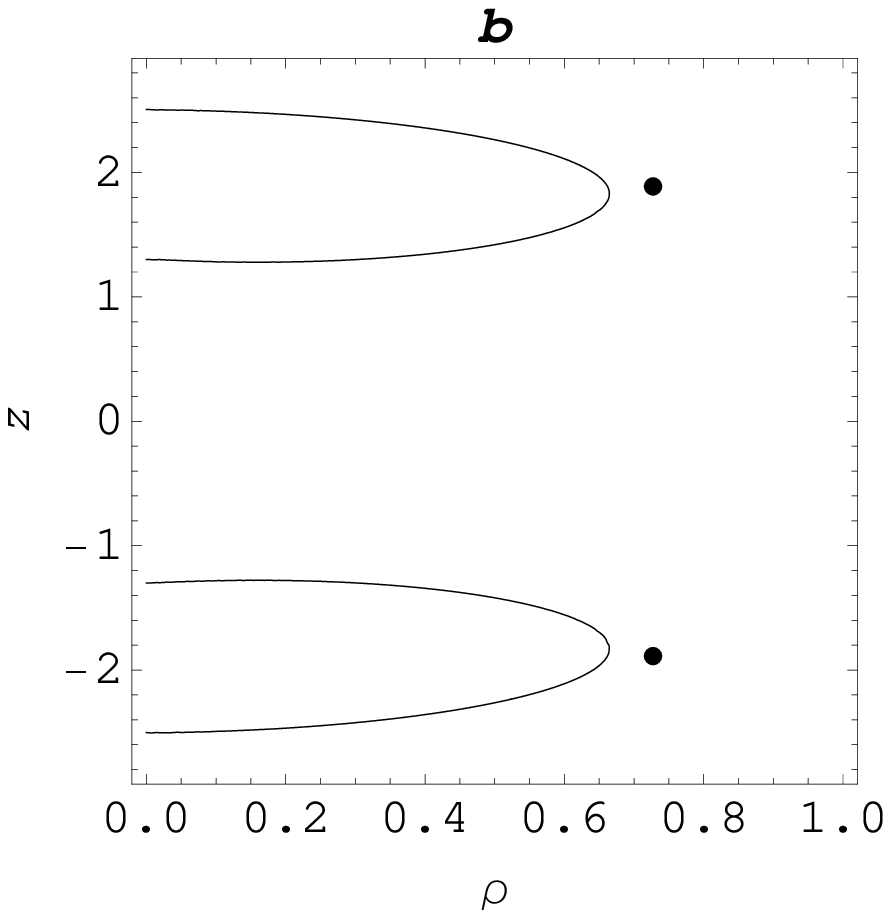}
\end{minipage}
\caption{(a) If $M>0$ there exist no singularities outside the ergosurface, and the ring singularity lies inside of it, on the symmetry axis, for the values $\sigma=0.6$, $Q_{E}=-0.3$, $Q_{B}=-0.4$, $R=3.813$, $M=1$, and $J=1.5$. (b) Emergence of ring singularities if $M<0$, for the values $\sigma=0.6$, $Q_{E}=-0.2$, $Q_{B}=-0.1$, $R=3.813$, $M=-1$, and $J=-0.345$. The singularities are located at $\rho\simeq 0.73$, $z\simeq \pm 1.89$.}
\label{SLS1}\end{figure}

In Fig. \ref{SLS2} we have plotted the stationary limit surfaces (SLS), for two identical counter-rotating extreme KN black holes, performed by setting $f=0$. Once again, the appearance of ring singularities off the axis is due to the presence of negative masses in the solution, Eq. (\ref{extreme}), and the electromagnetic charges moves the singularity outside the ergosurface.
\begin{figure}[ht]
\begin{minipage}{0.49\linewidth}
\centering
\includegraphics[width=4.25cm,height=5cm]{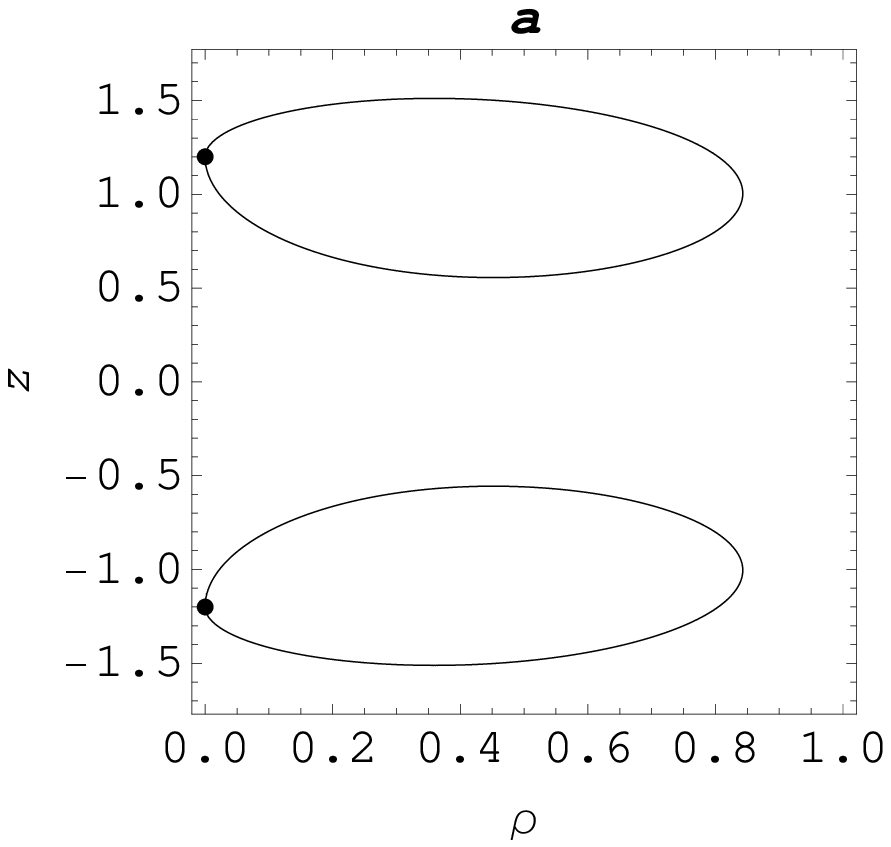}
\end{minipage}
\begin{minipage}{0.49\linewidth}
\centering
\includegraphics[width=4.25cm,height=5cm]{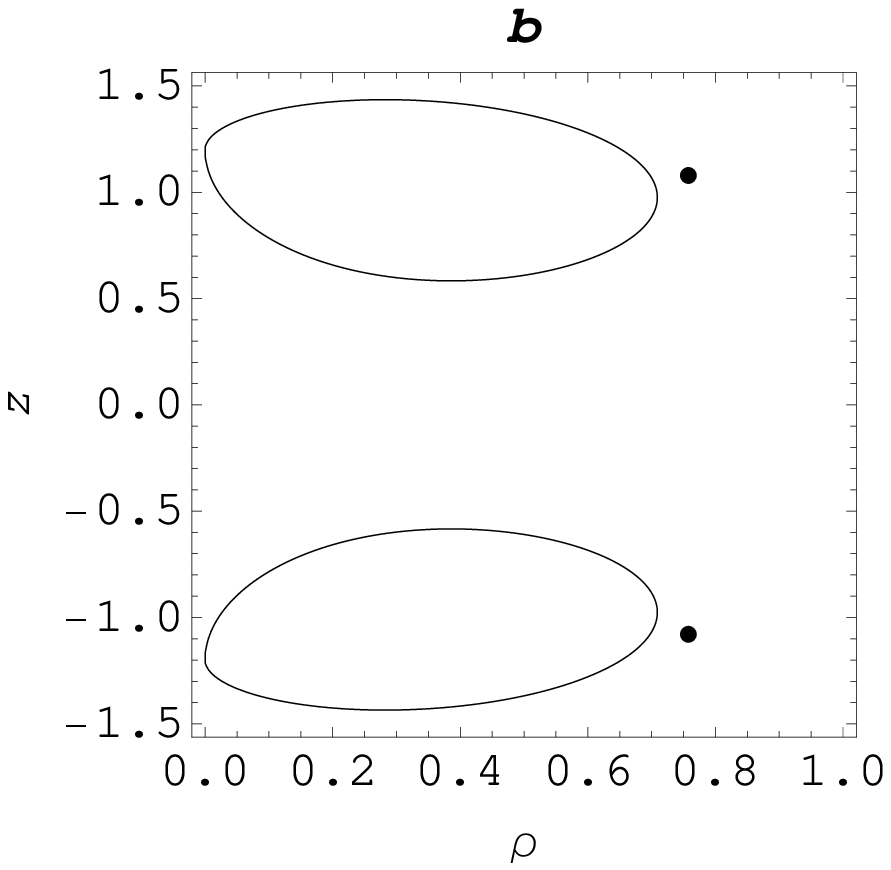}
\end{minipage}
\caption{(a) SLS for identical counter-rotating extreme KN black holes for the values $Q_{E}=-0.3$, $Q_{B}=0.2$, $R=2.4$, $M=1$, and $J=3.289$. (b) For the values $Q_{E}=-0.1$, $Q_{B}=0.1$, $R=2.4$, $M=-1$, and $J=-0.217$, the ring singularities are located at $\rho\simeq 0.76$, $z\simeq \pm 1.08$.}
\label{SLS2}\end{figure}

To conclude, recently some authors \cite{Manko} claimed that the Smarr formula does not suffer any change if one includes the magnetic charge into the solution. Nevertheless, this contradicts what Tomimatsu proposed \cite{Tomimatsu1} and Kleihaus \cite{Jutta} and Cabrera-Munguia \cite{ILLM} already confirmed. We have shown that the addition of the magnetic charge parameter $Q_{B}$ leads to a deeper understanding of the mathematical structure of this kind of spacetime. This result is quite naturally to be expected from a physical point of view. These authors \cite{Manko} do not like magnetic charges and complain about their unphysical nature, while they do not worry about the need for unphysical Weyl struts.

\section*{ACKNOWLEDGEMENTS}
We thank the referee for his valuable remarks and for checking the accuracy of our results.
This work was supported by CONACyT Grant No. 166041F3 and by a CONACyT fellowship with CVU No. 173252. C.L. acknowledges support by the DFG Research Training Group 1620 ``Models of Gravity'', and by the QUEST Center of Excellence.

\end{document}